\documentclass[english]{article}
\usepackage[T1]{fontenc}
\usepackage[latin9]{inputenc}
\usepackage{graphicx}
\usepackage{babel}
\usepackage{amsmath}

\makeatletter

\newcommand{\eq}[1]{Eq. (\ref{#1})}
\newcommand{\se}[1]{Sec. (\ref{#1})}
\newcommand{\la}[1]{ \label{#1}}
\renewcommand{\a}{\alpha}
\renewcommand{\b}{\beta}

\newcommand{\bsubs}{\begin{subequations}}
\newcommand{\esubs}{\end{subequations}}

\newcommand{\be}{\begin{equation}}
\newcommand{\ee}{\end{equation}}

\newcommand{\bea}{\begin{eqnarray}}
\newcommand{\eea}{\end{eqnarray}}

\begin{document}

\title{Expansions for Eigenfunction and Eigenvalues of large-$n$ Toeplitz Matrices \\
 }
\author{ Leo P. Kadanoff\\
The James Franck Institute\\
The University of Chicago}

\maketitle
\begin{abstract}

This note starts from work done by  Dai,  Geary, and Kadanoff\cite{DGK} on exact eigenfunctions for Toeplitz operators.  It builds methods for finding convergent expansions for eigenvectors and eigenvalues of large-$n$ Toeplitz matrices, using the infinite-$n$ case\cite{DGK} as a starting point.  One expansion is derived from operator equations having a two-dimensional continuous  spectrum of eigenvalues, which include the eigenvalues of the finite-$n$ matrices.  Another expansion is derived from the transpose equations, which have no  eigenvalues at all.  The two expansions work together to give an  apparently convergent expansion with an expansion parameter expressed as an inverse power of $n$.   A variational principle is developed which gives an approximate expression for determining eigenvalues.  A consistency condition is generated, which gives to lowest order exactly the same condition for the eigenvalue.

\end{abstract}
\tableofcontents


\section{Introduction}
\subsection{History}
This paper is a continuation of recent work by Dai, Geary, and Kadanoff\cite{DGK} (which we shall hereafter cite as paper I) and Lee, Dai and Bettleheim \cite{SYL+HD+EB} on the spectrum of eigenvalues and eigenfunctions for singular 
Toeplitz matrices.   A Toeplitz matrix is one  in which the marix elements, $T_{j,k}$, are functions of the difference between indices, $T(j-k)$.    To get sharp definitions of matrices and matrix elements of all orders, we define all matrix elements in terms of a single function, the symbol, $a(z)$ defined for $z=e^{-ip}$ on the unit circle.  Thus we write
\be
T_{j,k}= T(j-k)=\int \frac{dz}{2 \pi i z}  \frac{a(z)}{z^{j-k}}
\label{Toeplitz}
\ee

The basic problem under consideration here is the definition of a good method for calculating the eigenvalues and eigenfunctions of Toeplitz matrices, the matrices defined by letting the indices $j$ and $k$ live in the interval $[0,n-1]$. Previous authors\cite{Widom1,Widom2} have described the Toepltiz matrix problem by pointing out that the eigenvalues approach the spectrum of the analogous problem in which the indices vary over the set $[-\infty, \infty]$.  This latter problem may be solved by Fourier transformation and has an eigenfunction $\psi_j =e^{-ipj}$ and a corresponding eigenvalue $a(e^{-ip})$.  The set of all such eigenvalues, for real $p$, is terms the {\em image of the symbol}.   Widom speculates\cite{Widom1,Widom2} that in the large-$n$ limit, the discrete spectrum of the finite-$n$ problem approaches that image, at least for the case in which the symbol has a singularity on the unit circle.

Earlier authors\cite{SYL+HD+EB} established how this approach occurs for a very specific case in which the symbol had the form of singularity introduced by Fisher and Hartwig\cite{FH1,FH2} which is
\be
a(z)  =(2-z-1/z)^{\alpha}(-z)^{\beta}
\la{FH}
\ee 
Note that this singularity is defined by two parameters, $\a$, which defines  a zero in the symbol and, $\b$, which defines a discontinuity.  Using this symbol, the authors Lee, Dai, and Bettleheim\cite{SYL+HD+EB} found the spectrum for large $n$ and $\a=0$, while Dai, Geary and Kadanoff described a part of the spectrum for $0<\a<-\b<1$ in paper I.

The latter authors considered the behavior of Toeplitz operators.   These are Toeplitz matrices in which the indices run through the interval $[0,\infty]$.   Two cases should be differentiated in this analysis, when $0<\a<1$ and the image of the symbol forms a closed curve.:

case I . $0>\b>-1$.    All points within the image of the symbol are eigenvalues of the Toeplitz operator.\cite{DGK}   The finite-$n$ eigenvalues sit within that curve and approach it as $n$ goes to infinity.\cite{DGK}   

case II.  $0<\b<1$.  Again, the image of the symbol forms a closed curve. In fact, the curve depends only upon the absolute value of $/b$. But now, the Toeplitz operator has no eigenvalues.   However, the finite-$n$ eigenvalues of the corresponding finite-$n$ matrix  sit within that curve and approach it as $n$ goes to infinity.\cite{DGK}   

There is one more very interesting special case: $\a=0$.  In this situation, if $-1<\b<1$, the image of the symbol is a curved line segment, and the eigenvalues consist of all points which can be reached by connecting two points of that curve.   Once again the eigenvalues for finite-$n$ approach the curve while sitting within the region defined by the infinite-$n$ eigenvalues.\cite{SYL+HD+EB}

\subsection{The previous calculational strategy}
In the previous paper, paper I,  we studied the Toeplitz eigenvalue equation
\bsubs 
\be
\sum_{k=0}^{n-1}    T(j-k) \psi_k  = \epsilon {\psi}_j        \text{~~for~~}  0\leq j \leq n-1
\la{matrix}
\ee 
by studying a related Toeplitz operator equations
\be
\sum_{k=0}^{\infty}    T(j-k) {\psi^\infty}_k  = \epsilon {\psi^\infty}_j        \text{~~for~~}  0\leq j \leq \infty
\la{operator}
\ee
\esubs
for an eigenvalue for which both equations equally had solutions.  We could only solve the first equation numerically.   We had an exact method, the Weiner Hopf technique, for solving the second equation. 	
The crucial result was that the extension of the equation being solved  to the region between  $j=n$ and $j=\infty$ hardly changed the solution in the region of not-too-large values of $j$, specifically,  $0 \leq j \leq $  approximately $n/2$.  That is not too surprising since both equations have the same boundary behavior near $j=0$ and both may be expected to have the same asymptotic behavior for intermediate regions of $j$.   It would not be surprising, then, if we extended the analysis to include an equation which was right near $j=n-1$ and had the right asymptotics for smaller $j$, then we could get an asymptotic expansion which could work for all $j$.     In this paper, I precisely propose to extend the analysis by writing a pair of asymptotic statements, each one correct for one of the boundary regions.   

\subsection{plan of paper}
The next section includes the successive analysis based upon the two boundary regions.  The third section puts the two analyses together to get equations which will yeild an asymptotic expansion for eigenvalues and eigenfunctions.  The previous work, paper I, had a rather heuristic method for estimating the size of the corrections to the eigenvalue and eigenfunction.   Here we have an exact, testable expansion.  

\section{A pair of expansions}
\subsection{definitions}
In this section,  we look at  the Toeplitz matrix eigenvalue equation by using methods which work for solving the the Toeplitz operator eigenvalue equation, \eq{operator} .  In both cases, a solution is generated by extending the range of the index variables to $(-\infty,\infty)$, which will then permit us to use fourier transform techniques.  The equation for the Toeplitz matix's eigenvector can be 
in terms of three different kinds of functions which respectively are indicated by superscripts "-", "0", and "+".  The first superscript indicates a function which is non-zero only for $j<0$;  the superscript "0" defines a function non-zero for $0 \leq j  \leq n-1$; while the third superscript describes a function non-zero in $[n,\infty)$. The eigenfunction we wish to calculate is ${\psi^0}_j$ and it obeys 
 \begin{equation}
\sum_{k=-\infty}^{\infty}K_{j,k}{\psi^0}_k ={\phi^-}_{j} + {\phi^+}_{j}
\la{matrix1} 
\end{equation}
which now holds for all integer values of $j$. Here, the matrix $K$ is 
\be
K_{j,k} =T(j-k)-\epsilon \delta_{j,k}
\la{K1}
\ee

\eq{matrix1} will be analyzed in fourier transform language, with $z$ being the fourier variable.  Thus, the four quantities defined in that equation will be written as $K(z), \phi^-(z), \psi^0(z)$, and $\phi^+(z)$, which will respectively contain powers of $z$ extending from $-\infty$ to $\infty$;  only negative powers of $z$; non-negative powers extending up to $z^{n-1}$; and powers from $z^n$ to $z^\infty$. 
We also need to  define a  notation for the decomposition of the K operator.  We write
\be
K(z) = K^>(z) /(z K^<(z))
\la{K}
\ee
where $K^>$ has all its singularities and zeros outside the unit circle and $K^<$ has all its singularities and zeros inside the unit circle. These functions then respectively expand in a power series in $z$ and in $1/z$.     Correspondingly, the fourier transforms obey
$$
K^>(j-l) = 0     \text{~for~} j<l
$$       
while 
$$
K^<(j-l) = 0     \text{~for~}l<j
$$
Acting to the right, $K^>$ ($K^<$)  respectively carry information rightward toward higher $j$ values (leftward toward lower values).

\subsection{Weiner-Hopf  analysis for Toeplitz operator}   \la{WHA}

This section is not at all new.   It is all contained in paper I.  However, the notation is slightly different here.   We set $n=\infty$ and note that $\phi^+$ must be zero.  To distinguish the solution for the Toeplitz operator for the one for the Toeplitz matrix, we write $\Psi$ for the eigenfunction and $\Phi^-$ for the auxillary function $\phi^-$.  We then write the fourier transform of \eq{matrix} as
\be
K^> \Psi   = z K^<  \Phi^-
\la{WH}
\ee
Note that we have also dropped the superscript $0$ in our writing.  

\eq{WH} follows the usual Weiner-Hopf strategy.   The left hand side of the equation contains only non-negative powers of $z$.  The right hand side contains only non-positive powers.      Then both sides can only be constant, so that \eq{WH}  has the solution
\bsubs   \la{WHs}
\be
K^>  \Psi   =C
\la{WH1}
\ee
\be
z K^< \Phi^-  =C
\la{WH2}
\ee
\esubs
with $C$ being simply a constant in this fourier transform language. ( In coordinate space, $C$ becomes $C\delta(j,0)$.) 
The solution can then be written in terms of two functions:
\bsubs
\be
(1/K^>) (j)  =\int \frac{dz}{2 \pi i z} z^{-j}/K^>(z)           
\la{K>}
\ee
which vanishes for $j\leq 0 $   The other function is
\be
(1/K^<) (j)  =\int \frac{dz}{2 \pi i z} z^{-j}/K^<(z)            
\la{K<}
\ee
\esubs
The analysis in paper I enables us to describe the asymptotic structure of these functions for large values of $j$ in the previously analyzed case $0<\a<-\b<1$.   Their fourier transforms each contain a weak singularity  at $z=1$ proportional to $(1-z)^{2 \a}$.  This zero then produces a real term which decays as $1/j^{1+2\a}$.   In addition,   $K^>(z)$ has a simple zero at $z=z_c=e^{ip_c }$,  just outside the unit circle.   This result gives us an asymptotic form for large $j$
\bsubs
\be
\Psi^0_j=C (1/K^>) (j)    \sim  A^>/j^{1+2\a} + B e^{-ip_c j},          
\la{Psi}
\ee
This quantity vanishes for $j\geq 0 $ and, as we see,  decays algebraically  for large negative values of j. Paper I suggested that the two terms on the right-hand-side of \eq{Psi} were both of order $n^{-1-2\a}$ when $j$ is of order $n$.   This result gives us a small parameter  $\lambda \sim n^{-1-2\a}$, which might  be used in expansions.

The other  unknown functions can similarly be evaluated as
\be
\Phi^-\\_j  =  C  (1/K^<)(j-1)  \sim  A^<(-j)^{-1-2\a}
\la{WHO-}  
\ee
Here, the estimate holds when $j$ takes on large negative values.  When $j$ is greater than or equal to zero, this quantity is zero.   
\esubs

\subsection{Transpose of Toeplitz matrix}

There are two different forms of the equation for the Toeplitz matrix.   The first is 
 \eq{matrix}. 

The second form is a sort of transpose of \eq{matrix}, obtained by taking that equation in coordinate space and replacing each coordinate, $j$,  by  $j  \rightarrow \tilde{j} = n-1-j$.  We then get an equation in coordinate space 
$$
\sum_{k=0}^{n-1} K(n-1-j-k)  \psi^0\\_{k}  = \tilde{\phi}^-\\_j +\tilde{\phi}^+\\_j,
$$
where transposed quantities are defined as
$ \tilde{\phi}^\pm_j=  \phi^\mp_{n-1-j}  $  ,   $ \tilde{\psi}^0_j=  \psi^0_{n-1-j}  $.
and $\tilde{K}(j-k) =K(k-j)  $.    In this way, we find the second fundamental form for the eigenvalue equation
\be
\sum_{k=0}^{n-1} \tilde{K}(j-k)  \tilde{\psi}^0\\_k  = \tilde{\phi}^-\\_j +\tilde{\phi}^+\\_j
\ee
or, in Fourier transform, 
\be
 \tilde{K}(z)  \tilde{\psi}^0(z)  = \tilde{\phi}^-(z)+\tilde{\phi}^+(z)
\la{WHmet} 
\ee
where  the Fourier transform of the new Kernel is given by
\bea
\tilde{K}(z) &=&  K(1/z) =z  \tilde{K}^>(z)/\tilde{K}^<(z)   \nonumber\\
\tilde{K}^>(z) &=&1/K^<(1/z)      \nonumber \\
  \tilde{K}^<(z) &=&1/K^>(1/z)
\eea
and the new wave functions are given by
\bea
\tilde{\psi}^0(z)&=&z^{n-1}\psi^0(1/z) \nonumber  \\
\tilde{\phi}^\pm(z)&=&z^{n-1}\phi^\mp(1/z)
\eea
\subsection{Analysis for Toeplitz matrix: right eigenvector}
\eq{matrix}  can be analyzed using the same kinds of splitting of $K$ employed in \se{WHA}. We take that equation in its fourier transformed representation
$$
     K(z) \psi^0(z) =   \phi^+(z) + \phi^-(z)
$$
and multiply through by $z K^<$.  One then finds 
 \be
K^> \psi^0   = z K^<  \phi^- +z K^<  \phi^+
\la{WHme}
\ee
   The second term on the right hand side of this equation can be split up into parts which contain exponents of $z$ which are respectively negative, between zero and $n-1$ inclusive, and above n-1 in the form 
$$
z K^<  \phi^+= ( z K^<  \phi^+)^- + ( z K^<  \phi^+)^0 + ( z K^<  \phi^+)^+
$$
Then this equation can be rearranged into a form in which terms in non-positive powers of $z$ appear on one side and non-negative powers on the other. i.e.
 \be
K^> \psi^0- ( z K^<  \phi^+)^0 - ( z K^<  \phi^+)^+   = z K^<  \phi^- +(z K^<  \phi^+)^- =C
\la{WHm}
\ee
The second equal sign in this equation sets both sides equal to a constant, as in \eq{WHs}.

We solve for $\psi^0$, finding 
\be   \nonumber
 \psi^0= (1/K^> )C + (1/K^>)[( z K^<  \phi^+)^0+ ( z K^<  \phi^+)^+]
\ee
since $(1/K^>)$, acting to the right pushes coordinate indices toward higher  values,  
\bsubs
\la{direct}
\be
 \psi^0 =[ (1/K^> )C]^0 +[(1/K^>)( z K^<  \phi^+)^0]^0
\la{0}
\ee
The equation for the negative $j$ domain in \eq{WHm} may be solved for $\phi^-$ to give
\be
 \phi^- =(zK^<)^{-1}[ C  -  (z K^<  \phi^+)^-]
 \la{-}
\ee
\esubs
In contrast to the case of the Toeplitz operator, the equation for the eigenfunction is given in terms of the subsidiary functions $\phi^+$.   Further for the Toeplitz operator only the function $K^>$ is needed to determine the eigenfunction.  Here both $K^>$ and $K^<$ are involved.  These are important distinctions. 

Note the $1/z$ to the right of the equal sign in \eq{-}.    This factor has the effect of making the leading term in the expansion of $\psi^-$ be $1/z$, which is then followed by highter powers of $1/z$.  This is precisely the right structure for the expansion of $\phi^-$.

\eq{direct} gives us expressions for two of the quantities we need to know. However, we are far from done.   These equations give us relatively simple expressions for $ \psi^0$ and $\phi^- $, but we do not yet have an equivalently simple expression for $\phi^+$.     In both of the two subequations in \eq{direct},  we can evaluate the first term directly while  second term could be evaluated by quadratures if we but knew $\phi^+$.  Note that the lowest-order terms in both of these subequations are precisely the same as in the solution to the Toeplitz operator.

Our previous results\cite{DGK}, show that for small and intermediate values of $j$ in the set $[0,n-1]$, the first term in \eq{0} varies over a wide range, being of order $C$ for small values of $j$ and of order $C \lambda $, with 
\be
\lambda= 1/n^{2\a+1} \sim |e^{-ip_cn}|
\la{lam} \ee
for $j$ of order $n$. Similarly the first term in \eq{-} varies from being of order $C$ for $-j$ or order unity to being of order  $C \lambda $ for $-j$ of order $n$.   From the previous analysis one might guess that $\phi^+$ might be of the same order as the Weiner-Hopf values of $\psi_j/C$, for $j$ of order $n$, which is of order $\lambda$.   Thus, the correction terms in \eq{0} seems to be, at most, of this relative order while the correction term in \eq{-} appears to be of relative order $\lambda^2$.  So far, we have no expressions of corresponding simplicity for $\phi^+$ itself.   

\subsection{Analysis for Toeplitz matrix: left eigenvector}

It appears that  we have usable lowest order results for two of the three unknown functions.   The third unknown, $\phi^+$, contributes correction terms but it is hard to see a direct way to get it from \eq{WHm}.  We do have \eq{WHmet} in reserve from which we can get additional information. This equation has the consequence, 
\be
\tilde{K}^> \tilde{\psi}^0 = z^{-1} \tilde{K}^< \tilde{\phi}^+ + z^{-1} \tilde{K}^<  \tilde{\phi}^- 
\la{WHmet1}
\ee
We then split up the term  $ z^{-1} \tilde{K}^< \tilde{\phi}^+$  into  the different regions, $-$, $0$, and $+$ to derive a result analogous to \eq{WHm}, namely  
 \be
\tilde{K}^> \tilde{\psi}^0-( z^{-1} \tilde{K}^< \tilde{\phi}^+)^0 - ( z^{-1} \tilde{K}^<  \tilde{\phi}^+)^+   = z^{-1} \tilde{K}^<  \tilde{\phi}^- +(z^{-1} \tilde{K}^<  \tilde{\phi}^+)^-
\la{WHmt}
\ee
 There is, however, a substantial differences between \eq{WHm} and \eq{WHmt}.  The former has a $z$ in it, while the latter has a $z^{-1}$.  As a result, the first term to the right of the  equal sign in \eq{WHmt} cannot have a constant term in it.  It begins its expansion at $(1/z)^2$ and continues to higher powers of $1/z$.  The second term to the right of the equal sign also cannot have a constant term in it since $(. )^-$ is defined to start with a $1/z$ term.  Therefore the two sides of the equal sign cannot be equal to a non-zero constant.  They must both be zero.  This is, at bottom, a reflection of the fact that the transposed Weiner-Hopf operator equation has no eigenvalue solutions.
 
We thus see that \eq{WHmt} has the double consequence that  
\bsubs
\la{trans}
\be
 \tilde{K}^> \tilde{\psi}^0-( z^{-1} \tilde{K}^< \tilde{\phi}^+)^0 - ( z^{-1} \tilde{K}^<  \tilde{\phi}^+)^+  =0 
 \ee
 and 
 \be
  z^{-1} \tilde{K}^<  \tilde{\phi}^- +(z^{-1} \tilde{K}^<  \tilde{\phi}^+)^-=0
\la{-tp}
 \ee
 \esubs
We now write the transposed analogs of \eq{0} and \eq{-}, these analogs are \bsubs
\be
 \tilde{\psi}^0 =[(1/\tilde{K}^>)( z^{-1} \tilde{K}^<  \tilde{\phi}^+)^0]^0
\la{0t}
\ee
and 
\be
 \tilde{\phi}^- =  - (z/\tilde{K}^<) (z^{-1} \tilde{K}^<  \tilde{\phi}^+)^-
 \la{-t}
\ee
\esubs
These equations are simpler than \eq{0} and \eq{-}  because the terms in $C$ are lacking.  When the transposed quantities are replaced by their values in untransposed quantities, these equations read
\bsubs
\be
\psi^0 =[K^<( (z^{-1}/K^> ) \phi^-)^0]^0
\la{0p}
\ee
and 
\be
\phi^+ =  - z K^> [(z^{-1}/K^> )  \phi^-]^+
 \la{-p}
\ee
\esubs
\eq{0p} is another expression for the eigenfunction, analogous to \eq{0}.   We hope that the two equations are equivalent.   \eq{-p} gives us a usable expression for $\phi^+$, which can then be employed to give explicit values to the correction terms in \eq{0} and \eq{-}.   These four equations will give us the results we need for the various eigenfunctions.

\section{Results}

\subsection{Equations for eigenfunctions} 

We argue about the relative sizes of the various terms in the equations by saying that $1/K^>$  and $1/K^< $ serve as propagators which connect the regions described by the symbols $-$, $0$, and $+$.  Any connection between $-$ and $+$ is necessarily small,  as is any connection of $C$ to $+$. I assert that these connections are of order  $\lambda$.   This smallness makes terms involving several regions necessarily small and makes it possible for our expansions for eigenfunctions, given below,  to be rapidly convergent.  

One such equation is derived by substituting the expression for $\phi^-$ in \eq{-p}  which gives
\be
\phi^+ =  - z K^> [(z K^> K^<)^{-1}  C]^+  + z K^> \{(zK^>)^{-1} ) [ (z  K^<)^{-2}   \phi^+)]^-
\}^+
 \la{-pq}
\ee  
The argument which we have just gave makes the first term of order $\lambda$ and the second of order $\lambda^3$, and suggests that this expansions is likely to converge.
An analogous derivation gives us an equation for  $\phi^-$.   Start from \eq{-}  and substitute into that \eq{-p} for   $ \phi^+$, giving
\be
 \phi^- =(zK^<)^{-1}\{ C + (z^2 K^<   K^> [(z K^> )^{-1}  \phi^-]^+)^-\}
 \la{-pp}
\ee
For $j$ at or near $-1$ the first term in the curly brackets is of order $\lambda^0$ while the second is of order $\lambda^2$.   

Using the same method, we can find an equation for the eigenfunction
\be
 \psi^0 = [(1/K^> )C]^0 +[(1/K^>)( z K^<  \phi^+)^0]^0
\la{0ext}
\ee      

\subsection{Eigenvalue condition: an extremal}
It appears that we have a potentially convergent expansion for our eigenfunction. However, there is a potential difficulty.  The expansions cannot always converge.  The equation for the various functions determine an eigenvalue, and cannot possibly converge unless the eigenvalue condition is met.  

One conventional way of finding eigenvalues is through the use of an extremum. 

Consider the quantity:
\bea
Q[\chi]&=&  N/D     {~~with~~}   \nonumber \\
D  &=&      \sum _{l=0}^{n-1}  \tilde{\chi}_l ~ \chi_l  =\tilde{\chi}  \cdot \chi   \nonumber\\ 
N=& =& \sum_{j,k=0}^{n-1}  \tilde{ \chi}_j  \big[  K(j-k)-\epsilon \delta_{j,k}\big]\chi_k   = [\tilde{\chi} \cdot T \cdot \chi]
\la{extremal}
\eea
This quantity is zero when $\epsilon$ is an eigenvalue of the $n$-th order Toeplitz matrix  and $\chi$ is the corresponding  eigenfunction. If $\chi$ deviates from this eigenfunction by a small amount, then $Q$ is of order of the square of the deviation. Since the eigenvalues vary by an amount of order unity, we might expect that for an arbitrarily chosen smoothly varying $\chi$, $Q$ would be of order unity.   Of course, if $\epsilon$ is not an eigenvalue of the Toeplitz matrix, this extremal property is lost.

Now look at the special case in which $\chi$ is our lowest order approximation for the eigenfunction as given by the first term in \eq{0}, 
\be
\chi_j  =\Psi_j =(1/K^>)(j)      \text{~~for~~}   0\leq j\leq n-1  
\ee
(For simplicity, we have set $C=1$.)
With this choice the denominator has the value
\bsubs  \la{eestimate}
\be
D=\sum_{l=0}^{n-1} (1/K^>)(l) ~(1/K^>)(n-1-l)
\la{D}
\ee
The first step in simplifying the numerator is to replace the sum over $k=[0,n-1]$ by a sum over $k=[0,\infty]$  minus a sum over $k=[n,\infty]$.  The sum over $[0,\infty]$ vanishes so that
$$
N= - \sum_{j=0}^{n-1} \sum_{k=n}^{\infty}  \Psi_{n-1-j}~\Psi_{k} ~K(j-k)
$$ 
The $j$-sum is split into pieces. We sum over $j=[0,\infty]$ and subtract the piece $j=[n,\infty]$.  \eq{WH} then gives a result in which $N $  comes out as the sum of two terms, $N=N^{++} + N^{+-}$
which respectively have the values:
\be
 N^{++}=  \sum_{j,k=0}^\infty  \Psi_{j+n}~\Psi_{k+n}~K(-j-k-1-n)=\sum_{j,k=0}^\infty  (1/K^>)(j+n) ]~(1/K^>)(k+n) ~K(-j-k-1-n)
 \la{++}
\ee
and
\be    
N^{+-}= -\sum_{k=0}^\infty \Psi_{k+n}~\Phi_{-k-1}=-\sum_{k=0}^\infty  (1/K^>)(k+n) ~(1/K^<)(-k-1) 
\la{+-}
\ee
\esubs

The results of paper  I enable us to estimate the order of magnitude of the various terms in \eq{eestimate}.  The denominator of \eq{D}, has the magnitude $D=O( n \lambda)$, unless there is a cancellation in the sum.   The numerator term  $N^{++}$ has the order of magnitude $N^{++}=O(n \lambda^3)$ since each $K$ in the product is of order $n \lambda$ and the sum converges by  falling off algebraically.  The numerator term  $N^{+-}$ has the order of magnitude $N^{+-}=O( \lambda)$ since  $K^<$falls off quite rapidly from its values, of order 1, for small values of $k$. This estimate gives the maximum possible result. The value of this term may be much smaller, and even zero, if there is cancellation between the exponential and the algebraic terms in $(1/K^>)(k+n)$ for small values of $k$.  Indeed, I am expecting such a cancellation to occur precisely in the situation in which $\epsilon$ has been chosen to be an eigenvalue of the Toeplitz matrix.   In sum, our extremal, contains two terms. One term, $N^{+-}/D$, of order $1/n$, if there is no cancellation in \eq{+-} and the other of order  $\lambda^2$.  The latter is smaller, unless the, expected, cancellation occurs.  

Why should one expect the cancellation in \eq{+-}?  The variational quantity, $Q$ will approach zero quadratically as $\chi$ approaches the correct eigenfunction $\psi^0$.   In paper I,  it was shown that the normalized functions generated by the Toplitz operator differed from the ones for the Toeplitz matrix, in the case in which the eigenvalues are identical,  by an amount of order $\sqrt{\lambda /n}$.  Thus,  our extremal should be $o(  \lambda /n)$ if $\epsilon$ is an eigenvalue of the Toeplitz matrix. The fact that we need the cancellation in \eq{+-} to achieve the correct order of magnitude strongly suggests that the vanishing of $N^{+-}$  is an eigenvalue condition.  

\subsection{An apparently exact condition}

So far, our heuristics have suggested that an appropriate approximate eigenvalue condition is the vanishing of the quantity $N^{+-}$ in \eq{+-}.  Can we find a more exact eigenvalue condition? Where should we look?

In the Toeplitz-operator case, the explicit factor of $z$ in the expression, \eq{K}, for the kernel gives that kernel an eigenfunction with zero eigenvalue. Conversely the factor $z^{-1}$  in the transposed kernel gives it no eigenvalues at all. Consequently we should look to the most explicit appearance of $z^{-1}$ in a transposed equation to look for the possible failure of the existence of an eigenvalue. \eq{trans} are the transposed equations for the eigenfunction and subsidiary function.      The first place to look is in \eq{-tp}, which has the suggestive property that it contains an explicit factor of $z^{-1}$.  

Both sides of the equation are zero.  The left hand side is an expansion     in $1/z$ starting at $1/z$ and proceeding to higher order in $1/z$.  At each order,   the two terms on the left hand side of \eq{-tp} must cancel.     Specifically, because of the $z^{-1}$ in the first term, that term starts off with a possible non-zero term of order $z^{-2}$.   The same must be true for the other term on the left of \eq{-tp}.   Therefore, we must satisfy the condition
$$
(z^{-1}\tilde{K}^<\tilde{\phi}^+)_{-1}=0
$$
that the   constant ($z$-independent) term in $\tilde{K}^< \tilde{\phi}^+ $  must be zero.   As a consequence, in \eq{-p}, the $z^n$ or $j=n$ term in the square bracket must be zero.   In coordinate space:
\be
\sum_{k=-1}^{-\infty}  (1/K^>)(n-k-1)          \phi^-\\_k   = 0
\la{eigen0}
\ee     
This equation will determine the eigenvalues of the Toeplitz matrices.It is pleasing to notice that this equation is essentially the same as the result in the more approximate \eq{+-}.  The only difference is that the exact auxilary quantity, $\phi^{-}$ has replaced its approximation from the Toeplitz operator, $\Phi^{-}$.

To see the consequences of this eigenvector equation plug in the lowest order evaluation of $ \phi^-_k  $  from \eq{WHO-} to get
\be
\sum_{k=-1}^{-\infty}  (1/K^>)(n-k-1)     (1/K^<)(k)        = 0
\la{eigen}
\ee
This is our lowest order (in $\lambda$) expression for an eigenvalue condition.

\section{Looking forward}
We have now completed our task of constructing an analytic (albeit heuristic) structure for an eigenfunction expansion.  The work is plausible but not proven.   The next step might be to construct proofs of the convergence of these results or alternatively to back them up with good numerical work. I have not yet carried out these tasks.

The work, in fact, lacks two checks which one might hope to out into place using purely analytic means.   I have not checked that the analytics yields the Fisher and Hartwig\cite{FH1,FH2} results for the product of eigenvalues.   If fact, I cannot see from where  they might arise. I also have not checked that the direct and the transpose methods give exactly the same eigenfunction.   

Last, but certainly not least, I don't know how the eigenfunctions and eigenvalues at the ends of the spectrum might arise.  Numerical evidence shows that they are different from the results at the middle of the spectrum, but the difference has been left unexplored up to now. 

In some respects, it is very pleasing to see that there is yet room for good additional work on this problem.

\newpage{}

\end{document}